\documentclass{vgtc}                          

\ifpdf
  \pdfoutput=1\relax                   
  \usepackage{graphicx}                
  \DeclareGraphicsExtensions{.pdf,.png,.jpg,.jpeg} 
\else
  \usepackage{graphicx}                
  \DeclareGraphicsExtensions{.eps}     
\fi%

\graphicspath{{figures/}{pictures/}{images/}{./}} 

\usepackage{microtype}                 
\PassOptionsToPackage{warn}{textcomp}  
\usepackage{textcomp}                  
\usepackage{mathptmx}                  
\usepackage{times}                     
\usepackage{cite}                      
\usepackage{tabu}                      
\usepackage{booktabs}    
\usepackage[normalem]{ulem}

\newcommand{\change}[1]{{\leavevmode\color{black}#1}}

\onlineid{0}

\vgtccategory{Research}

\vgtcinsertpkg

\title{ConVIScope: Visual Analytics for Exploring Patient Conversations}

\author{
\vspace{-2mm}
Raymond Li\footnotemark[1] \hspace{2em}
Enamul Hoque\footnotemark[2] \hspace{2em}
Giuseppe Carenini\footnotemark[1] \hspace{2em}
Richard Lester\footnotemark[3] \hspace{2em}
Raymond Chau\footnotemark[3]
\\
\parbox{4in}{\scriptsize \centering 
\vspace{1em}
\thanks{e-mail: \{raymondl, carenini\}@cs.ubc.ca}~~Department of Computer Science, University of British Columbia \\
\thanks{e-mail: enamulh@yorku.ca}~~School of Information Technology,  York University\\
\thanks{e-mail: \{rlester, raychau\}@mail.ubc.ca}~~Department of Medicine, University of British Columbia}
}

\abstract{
The proliferation of text messaging for mobile health is generating a large amount of patient-doctor conversations that can be extremely valuable to health care professionals. We present ConVIScope, a visual text analytic system that tightly integrates interactive visualization with natural language processing in analyzing patient-doctor conversations. ConVIScope was developed in collaboration with healthcare professionals following a user-centered iterative design. Case studies with six domain experts suggest the potential utility of ConVIScope and reveal lessons for further developments.

} 

\CCScatlist{
  \CCScatTwelve{Human-centered computing}{Visu\-al\-iza\-tion}{Visualization design and evaluation methods}{}
}

\teaser{
  \centering
  \vspace{-3mm}
  \includegraphics[width=\linewidth]{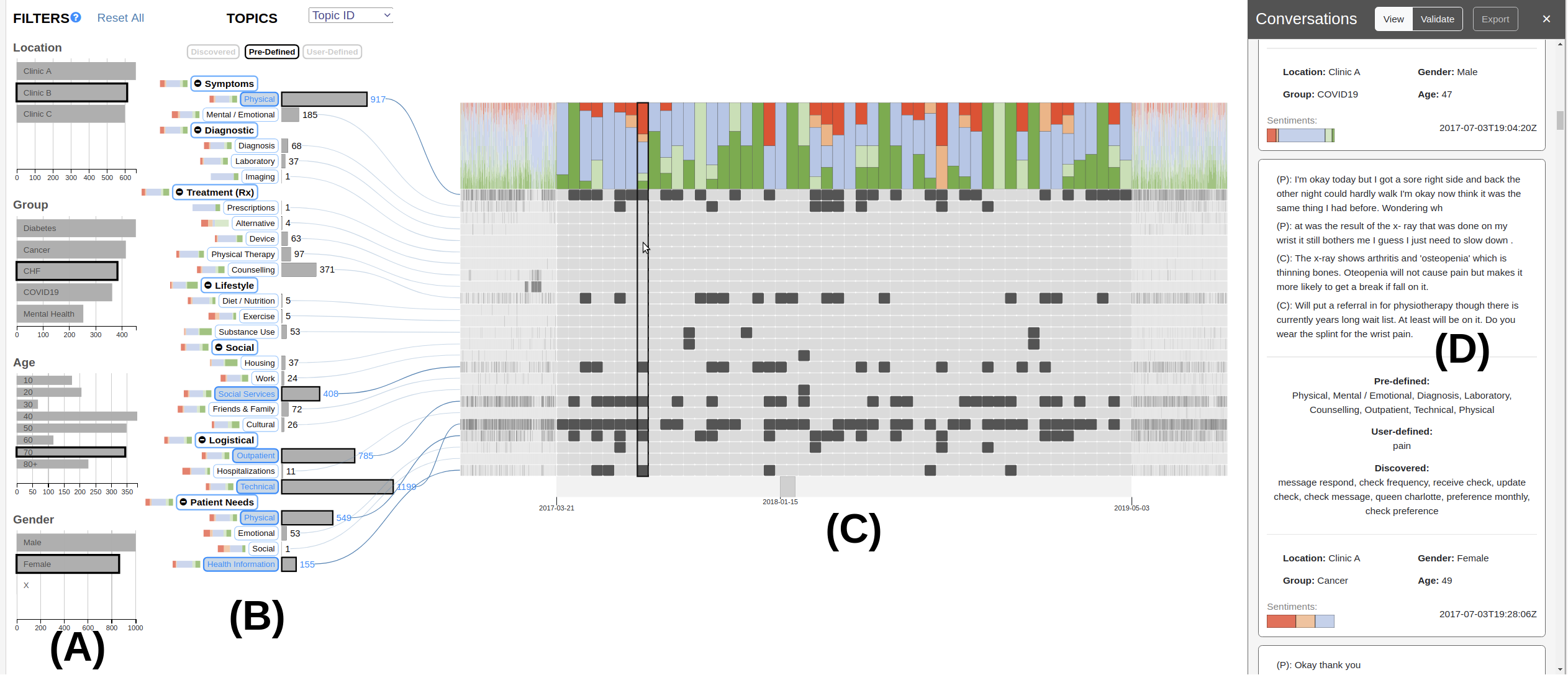}
  \vspace{-5mm}
  \caption{\textbf{ConVIScope} 
  Views (see \S\ref{sec:vis-design})
  : Metadata (A), Topics (B), Analysis (C) and  Conversations (D). 
  Trend View shown in \autoref{fig:trend}.
  }
  \vspace{-2mm}
  \label{fig:teaser}
}

\begin{document}


\vspace{-1mm}
\firstsection{Introduction}

\maketitle

Text messaging has become an important component of mobile health (mHealth) due to its wide adoption and affordable cost \cite{hall2015mobile}. As seen during the COVID-19 pandemic, physicians and health systems worldwide are shifting towards virtualized treatment programs to minimize physical visits and reduce hospital transmissions \cite{webster2020virtual}. In addition, studies involving texting-based mHealth programs have demonstrated their success at lifestyle modification \cite{haider2019mobile}, treatment adherence \cite{lester2010effects, ngwatu2018impact}, as well as disease monitoring \cite{kuehne2016mortality}.

The 
conversations generated 
from texting-based mHealth services can reveal valuable information about the patients
and the 
healthcare
quality.
For example, at the micro-level health professionals may be interested to quickly get an overview of what patients are discussing about, such as a concern with medications or logistical problems. At the macro-level, they may be interested to inform policy-making by knowing how conversations vary across different patient groups or to monitor what patients are complaining about in different regions.
Visual text analytic solutions could be effective in supporting the above tasks by helping clinicians to discover interesting trends and patterns among the patient population, and assist the health care administrators in optimizing the delivery of services. However, most existing 
solutions applied
Natural Language Processing (NLP) techniques on conversations extracted from online sources such as blogs \cite{hoque2014convis, hoque2016multiconvis}, forums \cite{fu2016visual, fu2018visforum}, and social media \cite{wu2014opinionflow, wu2017streamexplorer, so2020humane}. In contrast, 
visualizations that focus specifically to the healthcare domain primarily targeted 
healthcare records (e.g. \cite{plaisant1998lifelines, faiola2015supporting, kwon2018retainvis}) or online forums of health communities \cite{kwon2015visohc} as opposed to 
text messages between patients and doctors. 
\change{Finally, a few commercial mhealth platforms provide static charts for report generation but they do not support interactive visual exploration of a large set of messages~\cite{iribarren2017scoping}.}

In this paper, we conducted a design study with healthcare professionals to understand how to effectively combine NLP with visualization 
to support them in discovering \change{useful} trends and patterns from 
patient-doctor 
conversations 
and monitoring the delivery of health care services.
By following visualization design study methodologies~\cite{munzner2009nested, sedlmair2012design}, we first characterized the domain by interviewing experts to elicit their requirements.
We then performed data abstractions on real datasets obtained from one of our collaborators who is a healthcare researcher. 
After an iterative design process, 
we developed \textbf{ConVIScope}: a visual text analytic system that supports the exploration and analysis of patient-doctor conversations.
Our contributions include:
    \textbf{(1)} A requirement analysis from interviews with health 
    professionals, which revealed the unique challenges in understanding patient-doctor conversations and guided us to identify user tasks and key design needs.
    \textbf{(2)} The design and implementation of ConVIScope to meet the identified user requirements. 
    To the 
    best of our knowledge, ConVIScope is the first visual text analytics framework designed specifically for monitoring and exploring a large collection of patient-doctor conversations.
    \textbf{(3)} An evaluation 
    with six healthcare professionals which provides 
    initial evidence of the potential utility of ConVIScope and reveals lessons 
    for further developing visual 
    analytics systems for patient-doctor conversations.

\vspace{-1mm}
\section{Related Work}


\textbf{Visualization for Health Care:} Several early works on visualizing healthcare data mainly focused on electronic health records (EHR) data (e.g. \cite{plaisant1998lifelines, wang2008aligning, wongsuphasawat2011lifeflow, gotz2014methodology, chetta2015augmenting, glueck2017phenolines}). For example, Lifelines \cite{plaisant1998lifelines} organized individual patient records in expandable facets, while Lifelines2 \cite{wang2008aligning} provided a temporal summary of combined records. Lifeflow \cite{wongsuphasawat2011lifeflow} used icicle trees to organize records into hierarchical structures to reveal general patterns and trends. 
Others have  
 extracted and visualized text features from clinical data. For example, AdaptEHR \cite{hsu2012context} extracts medical concepts and maps them onto biomedical ontologies, while HARVEST \cite{hirsch2015harvest}  displays extracted concepts using a tag cloud and a timeline. RetainVis helps users interpret the prediction of neural models on medical records~\cite{kwon2018retainvis}. Unlike all the above works, which focus on summarizing and interpreting structured medical records, we focus on patient-doctor conversations.

\textbf{Visual Text Analytics for Conversations: } With the exponential growth of online conversations, 
there has been increasing 
interest in 
text analytics for such data. 
Several works focused on visualizing automatically extracted topics and sentiments from conversations~\cite{hoque2014convis,hoque2016multiconvis, el2016contovi, dou2013hierarchicaltopics}.
For example, ConVis \cite{hoque2014convis} shows the conversation structure and supports multi-faceted exploration using topics and authors. ConToVi helps to explore speaker behavior patterns in a  multi-party conversation using animations~\cite{el2016contovi}, while others have focused on organizing a large number of topics into hierarchical tree structures~\cite{dou2013hierarchicaltopics, hoque2016multiconvis}. Overall, although 
the above body of work targets multi-party conversations, they are not designed to deal with specific tasks of health professionals in exploring patient-doctor conversations, as we do in this paper. 

In this regard, 
work studying the visualization of medical conversations is very limited~\cite{lester2010effects, angus2012visualising, baker2015visualising}.
For example, Discursis~\cite{angus2011conceptual} 
has been applied to 
provide an overview of a patient-doctor conversation by visualizing the thematic contents as well as dynamics of turn-takings across speakers~\cite{angus2012visualising, baker2015visualising}.
However, these techniques mainly focus on the conversational structure while ignoring content facets, that we found to be critical in our study.
In contrast, VisOHC~\cite{kwon2015visohc} does visualize sentiments and topics of conversation threads, but from health forums, not from patient-doctor conversations. Also, it only focuses on supporting online health community administrators. Finally, MedStory \cite{sultanum2018more} aims to convey psychological and social aspects of a medical condition, by extracting and visualizing medical concepts and emotions. But again, its target are  social media conversations and not patient-doctor messages.  

\vspace{-1mm}
\section{Tasks and Data Abstractions}
\subsection{Requirement Analysis}
\label{sec:requirements}
In order to gather an initial set of requirements, we performed semi-structured interviews with eight healthcare professionals including 3 practicing physicians, 4 health researchers, and 1 healthcare administrator.
The interviews were open-ended in nature, where participants were asked about their needs for analysis and what they 
\change{need} to achieve with the insights derived from 
conversations with 
patients. 
In particular, the primary objectives of the healthcare professionals can be categorized into two scales. 
On a macro-scale, users require high-level insights across regions and demographics with the goal of designing more effective policies or supporting research objectives. 
On a micro-scale, users are interested in analyzing conversations from a particular patient group or clinic for identifying specific patient needs to improve the efficiency 
of the healthcare delivery.
Critically, the two scales are not orthogonal, as micro-level analysis can provide context for the macro-level and vice versa. 
In summary, we organized these goals as three generic requirements on both scales. First, the insights should allow the users to understand and compare the needs of the patients across different regions and demographics \textbf{(UR-1)}. Second, our interface 
should support 
evaluating the efficiency and quality of the healthcare services \textbf{(UR-2)}. Lastly, it is necessary for users to explore, identify and explain emerging issues and problems faced by the patients \textbf{(UR-3)}.

\vspace{-1mm}
\subsection{Task Model}

Based on the identified user requirements, we derived a set of analytical tasks that helps the domain experts in understanding and extracting insights from the patient-doctor conversations.



\textbf{T1. What is being discussed?}
Participants confirmed the importance of understanding the underlying topics being discussed in order to identify the patient needs (UR-1) and the emerging issues (UR-3) from the conversations. Often, they are interested in analyzing conversations with respect to 
predefined topics of medical relevance (e.g. treatment, prescriptions). Other times, they  are interested in exploring more specific and/or unforeseeable emerging topics beyond the pre-defined ones (e.g. mental 
problems from COVID-induced lockdown). To support these analytical needs, we extract
three 
types of topics 
(pre-defined, discovered, and user-defined). 



\textbf{T2. How are the attitudes being expressed?}
In addition to understanding what was discussed, users are also interested in how  
sentiments are expressed in  conversations, which  
can be a very useful indicator of patient satisfaction 
with the quality and efficiency of their healthcare service (UR-2). For example, the attitudes expressed by patients while discussing logistic issues (e.g., outpatient, hospitalization) can provide insights on  the efficiency of the healthcare service delivery. Moreover, visualizing the sentiment associated with a topic can help users to make policy decisions; for instance, addressing concerns about side-effects of a particular treatment.


\textbf{T3. What are the trends for topics and opinions?}
Understanding the trend associated with topics such as diseases and symptoms can help the users determine whether there are certain problems or events causing the prevalence of such topics 
(UR-3). For instance, the user might find an increasing number of conversations mentioning financial troubles due to the pandemics. Also, opinion trends can assist the user in assessing whether some aspects of the healthcare quality or patient satisfaction are changing on an aggregated level 
as well as deriving possible explanations for such changes (UR-2).



\textbf{T4. How do the conversations differ across different demographics and locations/clinics?}
Age plays an important role when considering phenotypic changes in health and disease \cite{geifman2013redefining}, and different patient groups (ex. cancer patients) may have different needs or face different problems (UR-1). Additionally, users are also 
interested in differences between conversations across clinics and regions, because they can reveal and help explain emerging 
trends occurring on a macro scale (UR-1, UR-3), providing insights on how to improve health policies 
that are tailored to those differences.
\vspace{-1mm}
\subsection{Data Model}



 \change{
 We pre-process the dataset by filtering out conversations containing less than 3 messages to focus on those that are long enough to perform meaningful studies.} 
 Each conversation is tagged with patient features such as location, patient group, age group, and gender. Additional metadata, including topics and sentiment, are derived from text content using the following NLP techniques.

\textbf{Pre-defined Topics:}
To satisfy the need of organizing conversations based on general medical knowledge (T1), 
in close collaboration with health professionals, we define a set of hierarchically organized topic labels (e.g. logistics, treatment, social - see 
~\autoref{fig:teaser} (B)). 
We refer to these labels as 
pre-defined topics. These topics 
are assigned to conversations by supervised classifiers trained on human-annotated data. We used logistic regression with bag-of-word features for this \change{task}  as it is a simple and interpretable classifier.  Moreover, it has shown promising results in recent work on predicting patient needs from social media conversations, with limited training data \cite{jang2019neural, lee2021identifying}. We involved healthcare professionals and medical students for annotating 
5K
conversations to train the classifiers. \change{To measure the inter-rater reliability, we asked four annotators to label the same set of $200$ conversations. The Cohen's Kappa coefficient was $0.64$, indicating a moderate level of agreement \cite{mchugh2012interrater}}.




\textbf{Machine-discovered Topics:}
In order to discover  
emerging topics beyond pre-defined topics (T1), we used the standard Latent Dirichlet allocation from the Gensim library \cite{rehurek2010software} for unsupervised topic model \cite{blei2003latent} \change{with $k=3$ as the number of topics.}
The label for each topic 
comprises the   five most likely words given that topic.

\textbf{User-defined Topics:}
Predefined topics cover 
generic medical terms like diagnosis, treatment, exercise etc. However, during an analysis, users may also become interested in more specific topics. ConVIScope allows the user to specify such topics on the fly enabling them to find conversations that contain a user-provided phrase. Such a mechanism also finds conversations with semantically similar phrases to the given one, based on cosine similarity between 
pre-trained embeddings for their words \cite{pennington2014glove}.

\textbf{Sentiment Analysis:}
To determine whether each message expresses a positive or negative attitude (T2), we use a lexicon-based method 
\cite{taboada2011lexicon}, which has been shown to deliver robust performance on conversational data~\cite{carenini2011methods}. 
We predict the sentiment value for each message which ranges from $-2$ to $+2$. 
\vspace{-1mm}
\section{Visualization Design}
\label{sec:vis-design}

In order to support the user requirements and tasks identified above, we iteratively designed 
ConVIScope 
over a 9 month period based on biweekly consultations with a team of health care researchers led by an medical specialist. 
These meetings provided 
invaluable feedback on refining and revising design choices.
The resulting final prototype (see \autoref{fig:teaser}) consists of multiple linked views including an overview 
summarizing the conversations with sentiment and topics using 
focus+context, 
and a detailed view 
showing the 
textual conversations. The system also supports filtering by multiple facets namely topics, time range, patient demographics, and locations. 

\vspace{-1mm}
\subsection{Visual Encodings}

The \textbf{Conversation Analysis View (C)}
represents a visual summary of the entire set of conversations based on topics and sentiments (T1, T2). It allows the user to explore through the corpus on a conversational level as well as identifying temporal trends and emerging patterns (T3).  
\change{Unlike ConVis \cite{hoque2014convis} which visualizes a conversation as an indented tree and connects the topics with tree nodes}, we use a \change{compact} heatmap \change{design} to encode the topic distributions, where each column represents a conversation, each row represents a topic (linked to the corresponding topic node in the topic hierarchy), and each cell indicates the  presence of a topic using dark grey color. At the top of each column, we use a vertical stacked bar to encode the sentiment distribution of the corresponding conversation where green indicates positive and red indicates negative sentiment.

\begin{figure}[tb]
 \centering
 \includegraphics[width=.95\columnwidth]{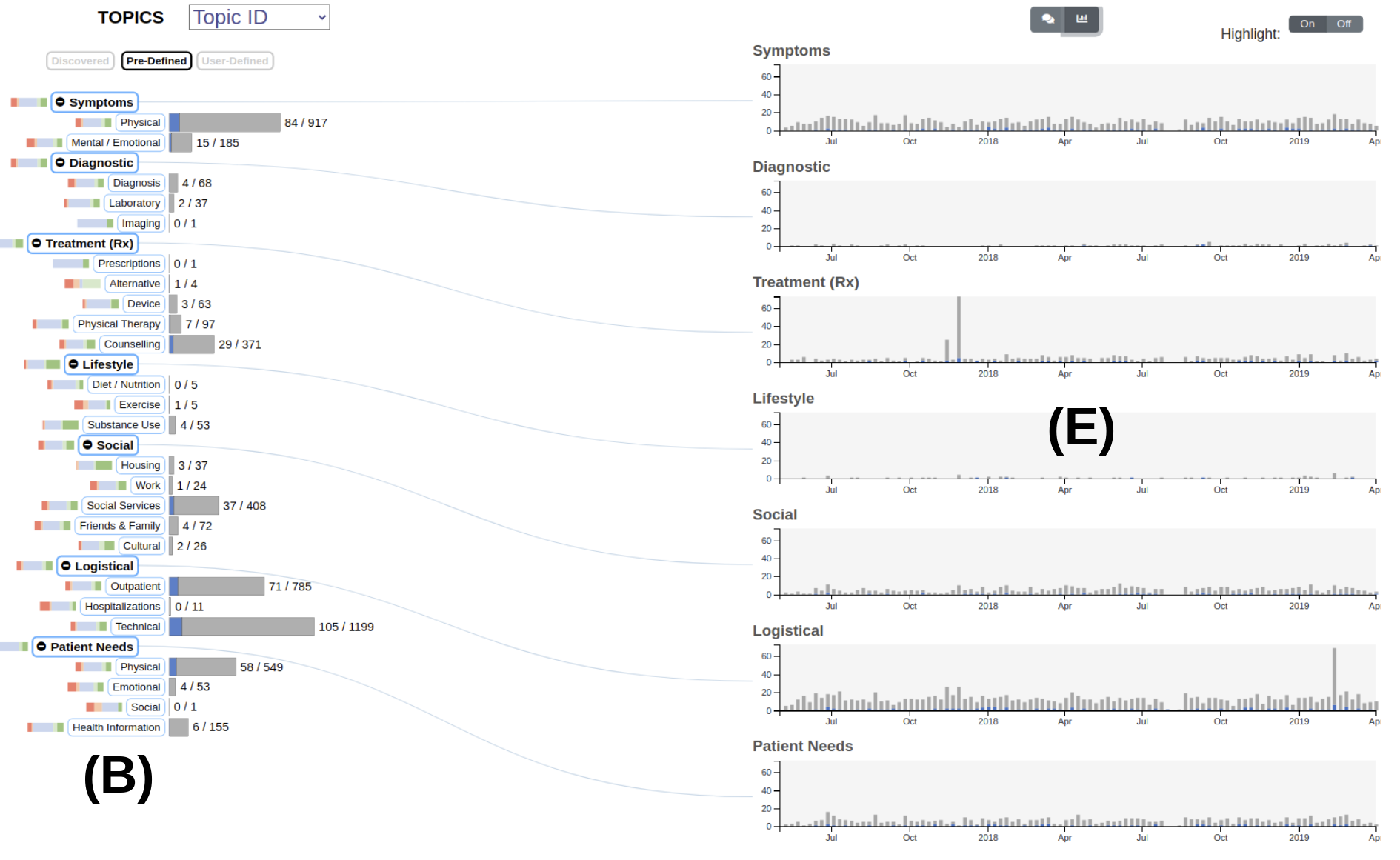}
 \vspace{-4mm}
 \caption{The Trend View (E) visualizes the number of conversations for each parent topic over each week.}
 \vspace{-5mm}
 \label{fig:trend}
\end{figure}

Overall, the Conversation Analysis View helps users identify how topics and sentiments are distributed across conversations and how they evolved over time in a space-efficient layout. With respect to MultiConVis \cite{hoque2016multiconvis} and VisOHC \cite{kwon2015visohc}, which can only show  a small number of conversations at a time, we 
improve scalability by designing a novel more compact representation that enables a focus+context approach. We opt for the focus+context technique to avoid the temporal (e.g. zoom) and spatial (e.g. overview+detail) separation by displaying the focus within the context in a single continuous display \cite{cockburn2009review}. 
The focused window visualizes a constant number of conversations within a fixed window, while the surrounding context 
(at the left and right side of this window)
summarizes the entire set of conversations by shrinking the resolution to fit all the columns inside the view. 
Finally, we include a separate \textbf{Trend View} that can be toggled to replace the Analysis View.
This Trend View allows the user to explore how the parent (more general) topics evolve over time, by showing a histogram  of the weekly volume of conversations  for each 
topic (\autoref{fig:trend} \change{E}).  

In the \textbf{Metadata (A)}
\textbf{and Topic View (B)}
we support the visualization of 
patient features and topics (T1, T4) by using frequency charts to convey the aggregated count for all the conversations containing the attribute. For patient features, we used a bar chart to visualize the number of conversations for each attribute (e.g. location, age, gender), where the attributes are the metadata tags from our data provider. Similarly,  the Topic View (B)
presents the list of topics where each topic is associated with a bar indicating the frequency of that topic. We also visualize the aggregate sentiment distribution of each topic by placing a horizontal stacked bar to the left side of a topic node to provide a summary of the attitudes from patients encountering specific problems (T2). Topics are organized in a hierarchy using indentation and are connected to the corresponding row in the heatmap via subtle curved links \cite{steinberger2011context}. In this way, the user can perceive which topics are being highly discussed (T1) and follow the associated rows in the Analysis View via the curved links.

The \textbf{Conversation View (D)}
is a scrollable list that displays the actual text of the conversation. 
The sentiment distribution, patient features as well as the timestamp for the first message in the conversation are displayed along with the text.

\subsection{User Interactions}

\textbf{Browsing through multiple granularity:}
The Analysis View initially provides an overview of all the conversations in the current dataset. Then, the user can create a focused window containing a constant number of conversations by  selecting a region in the scrollbar below the heatmap (C).
The user can move through the selection of a region in the scrollbar to change the focus which changes the context prior to and after the focused window. Clicking on any column representing a conversation causes scrolling to the relevant conversation in the Conversation View via a smooth animation.  This idiom allows the visualization to provide information at multiple granularities, \change{where the context 
represents the low-resolution summary, the focused window 
represents the high-resolution summary, and the most detailed view represents the actual conversation.}

\textbf{Faceted exploration:}
The user can perform faceted exploration by selecting attributes from the Metadata View and topic nodes in the Topic view. Hovering on any topic node results in highlighting the corresponding rows in the heatmap and the curved links that connect those rows with the hovered topic. Similarly, selecting a column in the heatmap results in highlighting relevant topic nodes in the Topic View for the corresponding conversation. Furthermore, features associated with that conversation are highlighted in the Metadata View by drawing a border around the corresponding rectangle in the bar charts. 
An example of this interaction is illustrated in \autoref{fig:teaser}, 
where the Metadata and Topic View respectively shows the corresponding patient features (`Clinic B', `CHF', `Age 70-80', `Female') and the topics (`Physical', `Social Services', `Outpatient', etc.). 

\begin{figure}[tb]
 \centering
 \includegraphics[width=\columnwidth]{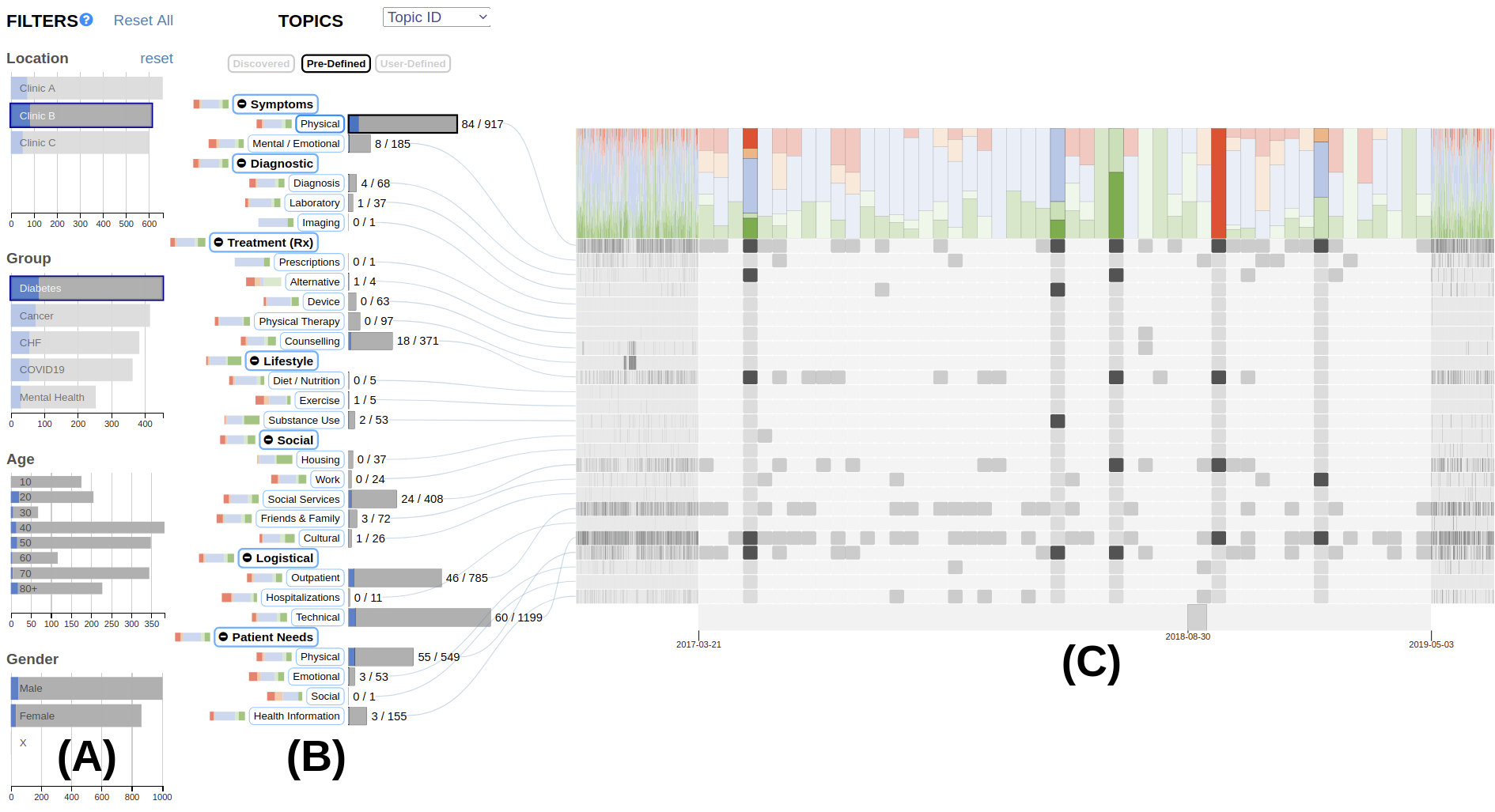}
 \vspace{-7mm}
 \caption{An example of  cross-filtering.}
 \vspace{-5mm}
 \label{fig:cross-filter}
\end{figure}

\textbf{Filtering via Selection:}
To support the user in selecting the conversations pertaining to specific patient features and topics, we employed a cross-filter approach. Specifically, whenever the user clicks on any item in the Metadata and Topic View  (e.g. a location, an age group) the interface shows the proportion of conversations containing that selected item across other patient features and topics. For example, when the user selects  `Female' under the `Gender' attribute, the proportion of conversations containing `Female' patients are highlighted using the blue colored component of the associated bar of each attribute value. Additionally, the columns representing conversations that do not match the selected attributes are filtered-out (de-emphasized) in the Analysis View.  \autoref{fig:cross-filter} demonstrates an example of cross-filtering
after the user selects three different criteria (``Diabetes patients with Physical Symptoms from Clinic B''). 



\change{\textbf{Interactive Labeling of 
Topics:}}
Since the predefined topics are classified \change{by the supervised model}, sometimes the predictions 
maybe inaccurate. 
To build more accurate models for predicting predefined topics (T1, T2), we include a feature that allows the user to interactively 
revise 
a pre-defined topic assignment based on their domain expertise. By enabling the `Validate' mode in the Conversation View (top (D)), the user can select whether they agree or disagree with the model prediction for a topic assignment based on the text content. At any time the user can  export the revised model
\change{in a CSV file} 
which can be used in future iterations of model training.

\vspace{-.4em}
\section{Expert Case studies}
To assess the efficacy of ConVIScope, we performed case studies with 6 domain experts. 
The goal was to understand: 
i) whether ConVIScope help users to perform the tasks identified in Section 3; 
ii) which visualization features worked and did
not work and 
iii) how to improve the system given the experts' feedback.


\textbf{Participants:} 
We ran the study with six domain experts (2 male, 4 female, age range 27-50 years). All participants were health professionals: 
$2$ practicing clinicians (C1, C2) and $4$ health researchers (R1-R4). Among them, C1, R1-R4 also participated in the requirement analysis interviews.
The participants held expertise in a variety of areas including infectious diseases, clinical psychiatry, public health, and physiology. The clinicians (C1, C2) were interested to use ConVIScope in improving the quality of practice as well as finding issues that patients are facing. Researchers (R1-R4) were more interested in various macro-aspects of patient conversations, such as studying the reasons for hospital admission and evaluating mHealth solutions for COVID-19 monitoring.

\textbf{Procedure:} The participants first went through short interviews, where we asked questions about their goals and the type of insights they hoped to get using ConVIScope. Then, they were provided with a 
\change{10-minute} tutorial about how ConVIScope works. Participants then accessed 
ConVIScope 
to explore a dataset consisting of 5775 conversations between 03/2017 to 04/2021 collected from a real-world healthcare facility by our collaborators. On average, each conversation had $4.5$ messages. 
The task was open-ended in nature where participants explored 
the dataset according to their own interests and they were free to use as much time as they needed. During the studies, we followed the think aloud protocol where we recorded all interactions and responses suggesting potential use cases and design feedback. 
The studies ended with semi-structured interviews, where participants answered questions about the usefulness of individual visualization components and the extent to which their goals were satisfied. All studies were conducted online, where participants shared their screens. \change{Each session lasted about an hour.}

\textbf{Interaction patterns:}
Reflecting the diversity of their goals and by leveraging the flexible interactions provided by  ConVIScope, participants displayed a very rich and diverse set of exploration strategies.
Half of them 
began their exploration using the Analysis View, while the other half began by interacting with the Metadata and Topics Views. 
In the Analysis View, participants mainly focused on the temporal trend of topics and sentiment by looking at the shrunken column in the surrounding context summaries, while using the focused window to select specific conversations for accessing their content in the Conversations View. 
 In the Metadata and Topic Views, 
 participants often used the cross-filter function, with researchers (e.g. R3 and R1) seemingly more interested in the aggregated counts of the selected conversations and correlations between topics and patient features (e.g. social issues for female cancer patients),
 while the two clinicians more frequently drilling down on the selected conversations in the Analysis View to determine whether particular topics were relevant for a patient group and then carefully reading  the associated conversations.
 
 

\textbf{Subjective feedback:} 
 Our analysis of the post-study interviews reveals that
 participants were impressed with ConVIScope. In particular, they highly appreciate the interface features 
for filtering conversations by topics and other facets (e.g. demographics and locations) and enabling them to verify 
what proportion of other facets belong to the filtered set of conversations through blue highlighting in the corresponding 
bar charts. For instance, R4 said,
\textit{``The number of (selected) conversations in the cross-filter can be used to supplement my research``}. Several participants also praised the heatmap component for providing an effective overview of the dataset:  \textit{``The (Analysis) View provides a good summary of the dataset, I can get a good idea of what's being talked about without reading the conversations``} (C1). The ability to drill down to the 
raw messages was also critical: \textit{``Conversation View is very important, where I verify whether the topics are actually being mentioned``} (R1). 

Admittedly, participants also raised some questions and suggested improvements. Multiple participants 
expressed uncertainty regarding the meaning of each sentiment bin,
\textit{``It will be nice to have the criteria for each sentiment color to see how the machine is thinking."} (R1), \textit{``What exactly is a neutral conversation?"} (R3). 
Questions were also raised about the support for comparison. For instance, R4 said:
\textit{``Although it's nice to compare conversations from two clinics by individually selecting them, it will be nice to distinguish between the two pooled results (compare the two selected clinics)}". 


\vspace{-3mm}
\section{Conclusion and Future Work}

We present ConVIScope, a visual analytic system that supports the exploration and analysis of 
patient-doctor conversations. 
Our system enables users to get an overview of a large set of conversations through multiple linked views, then filter through the set using topics and various metadata before 
narrow down to detailed messages. Results from our case study are encouraging, but multiple participants expressed the need for more interpretable sentiment analysis and for better support of comparison. These are the two main venues for future work. 
Further testing of the current prototype is ongoing at four different sites, where health care professionals can analyze their own data in a more ecologically sound setting. 

\vspace{-3mm}
\acknowledgments{
\vspace{-1mm}
We thank Edward Chiu, Will Choi, Jodi Gunawan, Chris Lee, and Abhishek Singh for their efforts in developing the interface. 
This work was supported 
by
Michael Smith Foundation for Health Research 
Award \#17273, with match-funding from WelTel Inc.}


\bibliographystyle{abbrv-doi}

\bibliography{references}
\end{document}